\documentclass[3p,times]{elsarticle}
\usepackage{epsfig}
\usepackage{ecrc}
\usepackage{amsmath}
\volume{00}
\firstpage{1}
\journalname{Physics Letters B}
\runauth{}
\jid{procs}
\jnltitlelogo{Physics Letters B}
\CopyrightLine{2022}{Published by Elsevier Ltd.}
\usepackage{amssymb}
\usepackage[figuresright]{rotating}
\begin{document}

\begin{frontmatter}

\title{The galaxy rotation curves in the $f(R,T)$ modified gravity formalism}


\author[label1]{H. Shabani}
\address[label1]{University of Sistan and Baluchestan, Faculty of Sciences, Department of Physics, Department of Physics, Zahedan, Iran\\
$^a$School of Astronomy, Institute for Research in Fundamental Sciences (IPM)
P. O. Box 19395-5531, Tehran, Iran}
\author[label2]{P.H.R.S. Moraes}
\address[label2]{Universidade Federal do ABC (UFABC) - Centro de Ci\^encias Naturais e Humanas (CCNH) - Avenida dos Estados 5001, 09210-580, Santo Andr\'e, SP, Brazil}

\begin{abstract}
Astronomical data have shown that the galaxy rotation curves are mostly flat in the far distance of the galactic cores, which reveals the insufficiency of our knowledges about how gravity works in these regimes. In this paper we introduce a resolution of this issue from the $f(R,T)$ modified gravity formalism perspective. By investigating two classes of models with separable (minimal coupling model) and inseparable (non-minimal coupling model) parts of the Ricci scalar $R$ and trace of the energy-momentum tensor $T$, we find that only in the latter models it is possible to attain flat galaxy rotation curves. 
\end{abstract}


\end{frontmatter}

\section{Introduction}
\label{sec:intro}

The dark matter paradigm is among the main challenges of experimental physics nowadays. Although we have a plethora of probes for dark matter gravitational effects \cite{schneider/1996}-\cite{amorisco/2022}, the search for dark matter particles have proved negative so far \cite{akerib/2017}-\cite{abdallah/2015}. The absence of detection of dark matter particles is referred to as {\it dark matter problem}. 

WMAP (Wilkinson Microwave Anisotropy Probe) observations of cosmic microwave background radiation temperature fluctuations point to $\Omega_c=0.235$ \cite{hinshaw/2013} as the density parameter of cold dark matter, which enhances the need for a deep understanding of dark matter physics. 

The same WMAP experiment points to $71.9\%$ of the universe composition to be in the form of the cosmological constant. The cosmological constant is what is expected to make the universe to accelerate its expansion, a phenomenon that is probed from measurements of the brightness from distant type Ia supernovae \cite{riess/1998}. 

Remarkably, the cosmological constant suffers from a fine tuning problem called {\it cosmological constant problem} \cite{weinberg/1989}. The astronomical observations, such as those in \cite{riess/1998}, indicate that the cosmological constant is many (up to $120$ orders of magnitude) smaller than estimations from particle physics \cite{weinberg/1989}. 

The cosmological constant problem has led to the construction of theoretical models attaining to explain the cosmic acceleration from different ingredients than the cosmological constant, such as quintessence \cite{tsujikawa/2013}-\cite{sahni/2002} and Chaplygin gas  \cite{chakraborty/2007,rudra/2013}. There are, actually, many other forms of describing the cosmic acceleration by replacing the cosmological constant with other ingredient (check, for instance, References \cite{tsyba/2011,lazkoz/2005}, among several others) and  the high degeneracy coming from this numerous possibilities is usually referred to as {\it dark energy problem}. 

The aforementioned problems can be approached through a different perspective, namely modified  theories of gravity. Those extend the General Relativity (GR) formalism in order to incorporate novel terms first in their actions and then in their field equations. Preferably, those new terms must be able to describe the dark matter and dark energy effects.

A good example of this possibility comes from the study of rotation curves of galaxies. Frigerio Martins and Salucci have used the $R^n$ gravity, which is a class of the $f(R)$ gravity models, for which $R$ is the Ricci scalar, with constant $n$, to well fit observations of rotation curves of galaxies with no need for dark matter \cite{frigerio_martins/2007}. Strong bonds were put to the $f(R)$ gravity through rotation curves of galaxies in \cite{vikram/2018}. However, solar system tests seem to rule out most of the $f(R)$ models proposed so far \cite{chiba/2003}-\cite{olmo/2007}. Also, relativistic stars cannot be presented in some $f(R)$ gravity models \cite{kobayashi/2008,goswami/2014}. Further $f(R)$ gravity shortcomings can be seen in Reference \cite{joras/2011}.

With the purpose of circumventing the shortcomings of the $f(R)$ gravity, Harko et al. have proposed to extend the formalism, by inserting terms dependent on the trace of the energy-momentum tensor $T$, giving rise to the $f(R,T)$ gravity \cite{harko/2011}. The dependence on $T$ in such a gravity theory may be due to quantum effects, imperfect fluids, extra fluids or an effective cosmological constant. The $f(R,T)$ gravity has been applied to several regimes, yielding interesting results, as one can check, for example, \cite{das/2016}-\cite{jamil/2012}. 

In what concerns the ``dark sector'' of the universe, $f(R,T)$ models able to accelerate the expansion of the universe were obtained in \cite{moraes/2016}-\cite{tiwari/2021}, for instance. The dark matter issue, on the other hand, was investigated in \cite{zaregonbadi/2016}. 

It is our purpose in the present letter to deeply investigate the rotation curves of galaxies in the trace of the energy-momentum tensor dependent modified gravity, namely $f(R,T)$ gravity. A dark matter halo is needed to fit observations of rotation curves of galaxies with newtonian dynamics predictions. Here, we will check if the $f(R,T)$ gravity is able to fit the observations in the galactic regime with no need to impose the existence of a dark matter halo. \\

\section{Trace of the energy-momentum tensor dependent modified gravity}\label{sec:temt}

Here we will highlight some important features of the $f(R,T)$ gravity formalism, following References \cite{harko/2011,shabani/2014}. We start from the $f(R,T)$ theory action, which reads

\begin{equation}\label{eq1}
S=\int \sqrt{-g} d^{4} x \left[\frac{1}{2\kappa} f(R,T)+L \right],
\end{equation} 
with $g$ being the metric determinant, $\kappa=8 \pi G$, $G$ is the newtonian gravitational constant, $f(R,T)$ is the generic function of $R$ and $T$, and $L$ is the matter Lagrangian density. Varying action (\ref{eq1}) with respect to the metric $g_{\mu\nu}$ yields the following

\begin{equation}\label{eq2}
F(R,T) R_{\mu \nu}-\frac{1}{2} f(R,T) g_{\mu \nu}+\Big{(} g_{\mu \nu}
\square -\triangledown_{\mu} \triangledown_{\nu}\Big{)}F(R,T)=\Big{[}8
\pi G+ {\mathcal F}(R,T)\Big{]} T_{\mu \nu}- {\mathcal F}(R,T)pg_{\mu \nu},
\end{equation}
as the $f(R,T)$ gravity field equations, with $F(R,T)\equiv\partial f/\partial R$, $R_{\mu\nu}$ being the Ricci tensor, ${\mathcal F}(R,T)\equiv\partial f/\partial T$ and $T_{\mu\nu}$ is the energy-momentum tensor. Moreover, we assumed $L=p$, with $p$ being the pressure. 

The covariant derivative of the energy-momentum tensor in the $f(R,T)$ gravitational theory can be written as the following

\begin{equation}\label{eq3}
\nabla^{\mu}T_{\mu \nu}=\frac{-\frac{1}{2}\mathcal {F}\nabla_{\mu}T
-T_{\mu \nu}\nabla^{\mu}\mathcal {F}+\nabla_{\nu}(p\mathcal{F})}{\kappa +\mathcal {F}}.
\end{equation}
Note, from Eq.~(\ref{eq3}), that, in principle, the $f(R,T)$ gravity does not predict the conservation of the energy-momentum tensor. In a cosmological perspective, this is related to a process of creation of matter through the universe evolution \cite{harko/2013}. 

In the present paper, we will consider models which respect the conservation of the energy-momentum tensor, i.e., models in which $\nabla^{\mu}{\sf T}_{\mu \nu}=0$. Assuming the equation of state $p=w\rho$, with constant $w$, the following constraints can be obtained from Eq.(\ref{eq3})

\begin{equation}\label{eq4}
\left\{
\begin{array}{l}
\frac{w-1}{2(w+1)}\mathcal {F}(R,T)=T\mathcal {F}_{T}(R,T)~~~~\mbox{minimal coupling models}\\
\left[\frac{w-1}{2}\mathcal {F}(R,T)-(w+1)T\mathcal {F}_{T}(R,T)\right]\dot{T}=(1+w)(3w-1)\dot{R}~~~~\mbox{non-minimal coupling models,}\\
\end{array}
\right.
\end{equation}
in which with {\it coupling} we mean {\it geometry-matter coupling}, that is, separable and non-separable parts of $R$ and $T$ in the $f(R,T)$ function. In (\ref{eq4}), a dot represents time derivative.

For the particular case of minimal coupling models $f(R,T)=R+h(T)$, in which $h(T)=\alpha T^{n}+\beta$, with $\alpha$, $n$ and $\beta$ being constants, the constraint~(\ref{eq4}) leads to $n=(3w+1)/[2(w+1)]$. 


By applying the Bianchi identity to the non-minimal coupling models $f(R,T)=R\tilde{h}(T)$, with $\tilde{h}(T)$ being a function of $T$, one obtains the lower constraint in (\ref{eq4}). In this case, the Ricci scalar is obtained by computing the trace of Eq.(\ref{eq2}). 

To discuss the problem of galaxy rotation curves we linearize the $f(R,T)$ gravity field equations (\ref{eq2}) using the Newtonian gauge through which the perturbed version of the Friedmann-Lem\^aitre-Robertson-Walker metric reads \cite{weinberg/2008}

\begin{equation}\label{eq6}
ds^{2}=-\Big{[}1+2\Psi(r)\Big{]}dt^{2}+a^{2}(t)\Big{[}1+2\Phi(r)\Big{]}\gamma_{ij},
\end{equation}
in which $\Psi(r)$ and $\Phi(r)$ are the metric perturbations, $a(t)$ is the scale factor and $\gamma_{ij}$ is the spatially flat three dimensional line element. 

Through this approach we consider a spherical mass embedding in the cosmic fluid. To proceed we perturb all scalar and tensor quantities included in the field equations (\ref{eq2}) as a sum of a time dependent spatially homogeneous background part and a time-independent one, i.e., $\Xi(t,r)=\Xi^{b}(t)+\Xi^{s}(r)$ (here the superscript $b$ stands for {\it background} while the superscript $s$ stands for {\it source}). Thus, using this assumptions we obtain 

\begin{align}\label{eq7}
&F^{b}R^{s}_{\mu\nu}+\left[F_R^b R_{\mu\nu}-\frac{F^{b}}{2}g_{\mu\nu}+\mathcal{F}_{R}^{b}\left(p^{b}g_{\mu\nu}-T^{b}_{\mu\nu}\right)+F_{R}^{b}\left(g_{\mu\nu}\Box-\nabla_{\mu}\nabla_{\nu}\right)+
\left(g_{\mu\nu}\Box-\nabla_{\mu}\nabla_{\nu}\right)
F_{R}^{b}\right]R^{s}=\nonumber\\
&\left(8\pi G+\mathcal{F}^{b}\right)T^{s}_{\mu\nu}+
\left[-F_{T}^{b}R^{b}_{\mu\nu}+\frac{\mathcal{F}^{b}}{2}g_{\mu\nu}+\mathcal{F}_{T}^{b}\left(T^{b}_{\mu\nu}-p^{b}g_{\mu\nu}\right)-F_{T}^{b}\left(g_{\mu\nu}\Box-\nabla_{\mu}\nabla_{\nu}\right)-(g_{\mu\nu}\Box
-\nabla_{\mu}\nabla_{\nu})F_{T}^{b}
\right]T^{s}-\mathcal{F}^{b}p^{s}g_{\mu\nu},
\end{align}

In the next sections we will investigate solutions of (\ref{eq7}) for a spherical mass distribution to model galaxy rotation curves. Two different types of $f(R,T)$ functions will be considered.

\section{Galaxy rotation curves in the minimal coupling $f(R,T)$ gravity}

In a static case such as the hydrostatic stellar equilibrium or rotation curves of galaxies, it might be valuable to search for ``conservative'' cases of the $f(R,T)$ formalism (check \cite{dos_santos/2019} for an example of the former case). Using Eq.~(\ref{eq2}), the equations of motion for the background cosmic matter read

\begin{align}
&3H^{2}=\kappa\rho^{b}+\alpha\frac{2n-w+1}{2(3w-1)}[(3w-1)\rho^{b}]^{n},\label{eq8}\\
&-2\dot{H}-3H^{2}=\kappa w\rho^{b}+\frac{\alpha}{2}[(3w-1)\rho^{b}]^{n},\label{eq9}
\end{align}
in which $H$ is the Hubble parameter ($\rho^{b}$ and $p^{b}$ denote the time-dependent parts of the density and the pressure of cosmic matter, respectively). Also, we consider a perfect fluid with $p^{b}=w\rho^{b}$ equation of state. To avoid ambiguity in the brackets one has to use $T\to-T$ for matter fluids with negative trace of the energy-momentum tensor. 

The potentials $\Psi$ and $\Phi$ are obtained by solving Eq.~(\ref{eq7}), which gives 

\begin{align}
&-\frac{2}{a^{2}(t)}\nabla^{2}\Phi(r)=\eta^{b}_{\rho}\rho(r)+\eta^{b}_{p}p(r),\label{eq10}\\
&\frac{2}{ra^{2}(t)}\left[\Psi'(r)+\Phi'(r)\right]=\zeta^{b}_{\rho}\rho(r)+\zeta^{b}_{p}p(r),\label{eq11}\\
&\frac{1}{a^{2}(t)}\nabla^{2}\left[\Psi(r)+\Phi(r)\right]-\frac{1}{ra^{2}(t)}\left[\Psi'(r)+\Phi'(r)\right]=\zeta^{b}_{\rho}\rho(r)+\zeta^{b}_{p}p(r),\label{eq12}
\end{align}
with 
\begin{align}\label{eq13}
&\eta^{b}_{\rho}=\kappa+\alpha n j\frac{4w-1}{3w-1}\left[(3w-1)j\rho^{b}\right]^{n-1},~~~~~~~
\eta^{b}_{p}=-\frac{\alpha nj}{4}\frac{3w+1}{3w-1}\left[(3w-1)j\rho^{b}\right]^{n-1},\\
&\zeta^{b}_{\rho}=-\frac{\alpha nj}{2}\left[(3w-1)j\rho^{b}\right]^{n-1},~~~~~~~~~~~~~~~~~~
\zeta^{b}_{p}=\kappa+\frac{3\alpha nj}{2}\left[(3w-1)j\rho^{b}\right]^{n-1},\nonumber
\end{align}
where
\begin{equation}\label{eq14}
j=\left\{
\begin{array}{l}
-1~~~~~~~~T<0~~ or~~ w<\frac{1}{3},\\
+1~~~~~~~~T>0~~ or~~ w>\frac{1}{3},\\
\end{array}
\right.
\end{equation}
(in Eqs.~(\ref{eq10})-(\ref{eq12}), $\rho(r)$ and $p(r)$ indicate the density and the pressure of a spherical configuration of matter which is immersed in the background cosmic matter). The superscript ``s" has been dropped for the time-independent variables, for abbreviation. In definitions~(\ref{eq13}) one obtains $\rho^{b}$ from Eq.~(\ref{eq8}), which for a static solution (and also the scale factor $a(t)$) has to be calculated in a fixed cosmic time. Assuming the present time, for $w=0$, one reaches the solutions $\rho^{b}_{0}=[\alpha/(1-\Omega_{0})]^{2}$ and $\rho^{b}_{0}=[\alpha\Omega_{0}/(1-\Omega_{0})]^{2}$ (where $\Omega$ denotes the baryonic matter density parameter and the subscript ``$0$" denotes present time values)\footnote{Easily, two solutions for the matter density can be understood from Eq.~(\ref{eq8}) for $w=0$. Note that, from the discussion following Eq.(\ref{eq4}), one gains $n=1/2$ for $w=0$.}. Therefore, one can calculate the coefficients~(\ref{eq13}).  Equation~(\ref{eq9}) simply gives the potential $\Phi(r)$ and by a simple manipulation of Eqs.(\ref{eq10})-(\ref{eq12}) one obtains
\begin{align}\label{eq15}
\frac{2}{a^{2}(t)}\nabla^{2}\Psi_{min}(r)=\left[3\zeta^{b}_{\rho}+\eta^{b}_{\rho}\right]\rho(r)+\left[3\zeta^{b}_{p}+\eta^{b}_{p}\right]p(r),
\end{align}
in which the subscript ``min" stands for minimal coupling models. Therefore, in the minimal coupling $f(R,T)$ gravity theories, the Newtonian gravitational potential admits a correction term, as $\Psi_{min}(r)=\Psi_N(r)+\Psi_c(r)$, which depends on the properties of cosmic matter. Obviously, this correction term depends on the inverse of distance of observation points like the Newtonian potential. Thus, the flatness problem of galaxy rotation curves cannot be solved in this context. From the the r.h.s of Eq.(\ref{eq7}) we see that for those models with $F_{T}^{b}\neq0$ there appears a new term corresponding to the derivatives of $T^{s}$ which sources our motivation to consider a non-minimal model in the subsequent section.

\section{Galaxy rotation curves in the non-minimal coupling $f(R,T)$ gravity}

In this section we obtain the solution of the perturbed field equation (\ref{eq13}) for the non-minimal coupling case, namely $f(R,T)=R[1+\beta h(T)]$ with constant $\beta$ and  $h(T)=T^{n}$, with constant $n$. In this case, the background equations of motion read

\begin{align}
&\left\{\beta\left[\frac{n(w+1)(9w+1)}{3w-1}-1\right]\left[(3w-1)\rho^{b}\right]^{n}-1\right\}H^{2}+2n\beta\frac{w+1}{3w-1}\left[(3w-1)\rho^{b}\right]^{n}\dot{H}=-\frac{\kappa}{3}\rho,\label{eq16}\\
&3\Bigg\{-1+\beta\Big[n (w+1)-1\Big] \Big[3 n (w+1)-1\Big]\Bigg\}\left[(3w-1)\rho^{b}\right]^{n}H^{2}+
\Bigg\{-2+\beta\Big[-2+3n(1+w)\Big]\left[(3w-1)\rho^{b}\right]^{n}\Bigg\}\dot{H}=\kappa w\rho.\label{eq17}
\end{align}

One can check that Eqs.(\ref{eq16}) and (\ref{eq17}) reduce to the usual GR form for $\beta=0$. Therefore, the time dependent background variables are supposed to be obtained from solving the lower relation in the results~(\ref{eq4}), Eqs.(\ref{eq16}) and (\ref{eq17}) for the chosen model. The perturbed equations for a static spherical mass read

\begin{align}
&2\left\{1+\beta\left[h^{b}-2(w+1)\rho^{b}h'^{b}\right]\right\}\nabla^{2}\Phi(r)-2\beta(w+1)h'^{b}\rho^{b}\nabla^{2}D(r)=-\kappa\rho(r)-\beta h'^{b}\nabla^{2}\left[-\rho(r)+3p(r)\right],\label{eq18}\\
&2(1+\beta h^{b})\frac{D'(r)}{r}=\kappa p(r)-\frac{2\beta h'^{b}}{r}\left[-\rho'(r)+3p'(r)\right],\label{eq19}\\
&(1+\beta h^{b})\left[\nabla^{2}D(r)-\frac{D'(r)}{r}\right]=\kappa p(r)+\beta h'^{b}\left(-\nabla^{2}+\frac{d}{rdr}\right)\left[-\rho(r)+3p(r)\right],\label{eq20}
\end{align}
in which the prime denotes a derivative with respect to the argument, $D(r)\equiv\Psi(r)+\Phi(r)$ and $a_{0}=1$ has been used. Also, the argument $T$ has been dropped from the corresponding functions\footnote{Obviously, we have $h'^b(\rho)=(3w-1)h'^b(T)$.}. Again, by setting $\beta=0$ one gets the corresponding GR equations. 

The system of equations (\ref{eq18})-(\ref{eq20}) yields the results
\begin{align}
&\nabla^{2}\Phi(r)=\frac{1}{2\mathcal{A}^{b}}\left\{-\kappa\rho(r)+\kappa\mathcal{B}^{b} p(r)+\mathcal{C}^{b}\nabla^{2}\left[-\rho(r)+3p(r)\right]\right\},\label{eq21}\\
&\nabla^{2}\Psi(r)=\kappa\mathcal{D}^{b}\rho(r)+\kappa\mathcal{E}^{b}p(r)-\mathcal{M}^{b}\nabla^{2}\left[-\rho(r)+3p(r)\label{eq22}\right],\\
&\mathcal{A}^{b}=1+\beta\left[h^{b}-2(1+w)h'^{b}\rho^{b}\right]\nonumber,~~~~~~~~~~\mathcal{B}^{b}=3(w+1)\beta\frac{h'^{b}}{1+\beta h^{b}}\nonumber,\\
&\mathcal{C}^{b}=-\frac{\beta h'^{b}}{1+\beta h^{b}}\left[2(w+1)\beta h'^{b}\rho^{b}+1\right],~~~~\mathcal{D}^{b}=\frac{1}{2\mathcal{A}^{b}},\nonumber\\
&\mathcal{E}^{b}=\frac{3}{1+\beta h^{b}}\left[\frac{1}{2}-(1+w)\beta\rho^{b}h'^{b}\right],~~~~~~
\mathcal{M}^{b}=\frac{\beta h'^{b}}{2(1+\beta h)\mathcal{A}^{b}}\left\{1+\beta\left[2h-6(w+1)h'^{b}\rho^{b}\right]\right\}.\nonumber
\end{align}

Comparing Eqs.(\ref{eq10}) and (\ref{eq15}) with (\ref{eq21})-(\ref{eq22}) one finds an extra term in the r.h.s of the latter which is proportional to the Laplacian of the matter density and pressure. To show how Eqs.~(\ref{eq21})-(\ref{eq22}) describe the galaxy rotation curve we consider a pressureless fluid permeating the galaxy, that is we assume $p(r)=0$. Note that because of different coefficients in Eqs.(\ref{eq21}) and (\ref{eq22}) one does not expect the same solution for the two potentials as in GR. In the absence of  pressure, the modified Newtonian potential obeys the following equation

\begin{align}\label{eq23}
\nabla^{2}\Psi_{nmin}(r)=\kappa\mathcal{D}^{b}\rho(r)+\mathcal{M}^{b}\nabla^{2}\rho(r),
\end{align}
in which the subscript ``nmin" stands for non-minimal coupling models. To continue, we firstly solve the above equation for known potential functions. In the limit of large distances from the center of the galaxy, astronomical data mostly predict a flat rotational curve, i.e., $v^{far}_{grc}=\sqrt{\mathcal{V}}=constant$ (in which the subscript ``grc" denotes ``galaxy rotation curve"), which means we are facing a gravitational potential of the form $\Psi(r)=\mathcal{U}+\mathcal{V}\log(r)$, for constant $\mathcal{U}$. Solving Eq.(\ref{eq23}) for this long distance potential we obtain

\begin{align}\label{eq24}
\rho^{far}(r)=\frac{1}{2\mathcal{M}^b r}\left\{2\mathcal{M}^b \mathfrak{c} e^{ -r \sqrt{-\frac{\mathcal{D}^b }{\mathcal{M}^b }}}+\mathcal{V}\sqrt{-\frac{\mathcal{M}^b}{\mathcal{D}^b}}\left[e^{ r \sqrt{-\frac{\mathcal{D}^b }{\mathcal{M}^b }}}Ei\left(- \sqrt{-\frac{\mathcal{D}^b }{\mathcal{M}^b }}r\right)-e^{- r \sqrt{-\frac{\mathcal{D}^b }{\mathcal{M}^b }}}Ei\left( \sqrt{-\frac{\mathcal{D}^b }{\mathcal{M}^b }}r\right)\right]\right\},
\end{align}
in which $Ei(x)$ is the Exponential Integral Function. The solution (\ref{eq24}) has been obtained from the condition that $\rho(\infty)=0$, with $\mathfrak{c}$ being an integration constant. The above solution behaves like $1/r$ in far distances for appropriate values of  $\beta$ and $w$. 

Near the center of a galaxy we take the approximation that the velocity rotation curves vary as $v^{near}_{grc}=\mathfrak{D}r$ for arbitrary values of the constant $\mathfrak{D}$. In this case, the solution of Eq.(\ref{eq23}) is given by

\begin{align}\label{eq25}
\rho^{near}(r)=\frac{1}{r\mathcal{D}^b}\left[2\mathfrak{D}\left(-1+\cosh\sqrt{\frac{\mathcal{D}^b }{\mathcal{M}^b }}r\right)+\sqrt{\mathcal{D}^b\mathcal{M}^b}\rho_{0}\sinh\sqrt{\frac{\mathcal{D}^b }{\mathcal{M}^b }}r\right],
\end{align}
in which $\rho_{0}$ is the galaxy density at the center. 

Fig.~\ref{fig1} below shows the behavior of (\ref{eq24})-(\ref{eq25}) solutions for the same values of $\beta$ and $w$ parameters as well as $n=1/2$. It can be seen that the non-minimal coupling $f(R,T)$ models are capable of justifying a matter density profile with an admissible behavior while the source and the background pressures are vanished, i.e., $w\approx 0$ and $p(r)=0$.

\begin{figure}[h!]
\begin{center}
\epsfig{figure=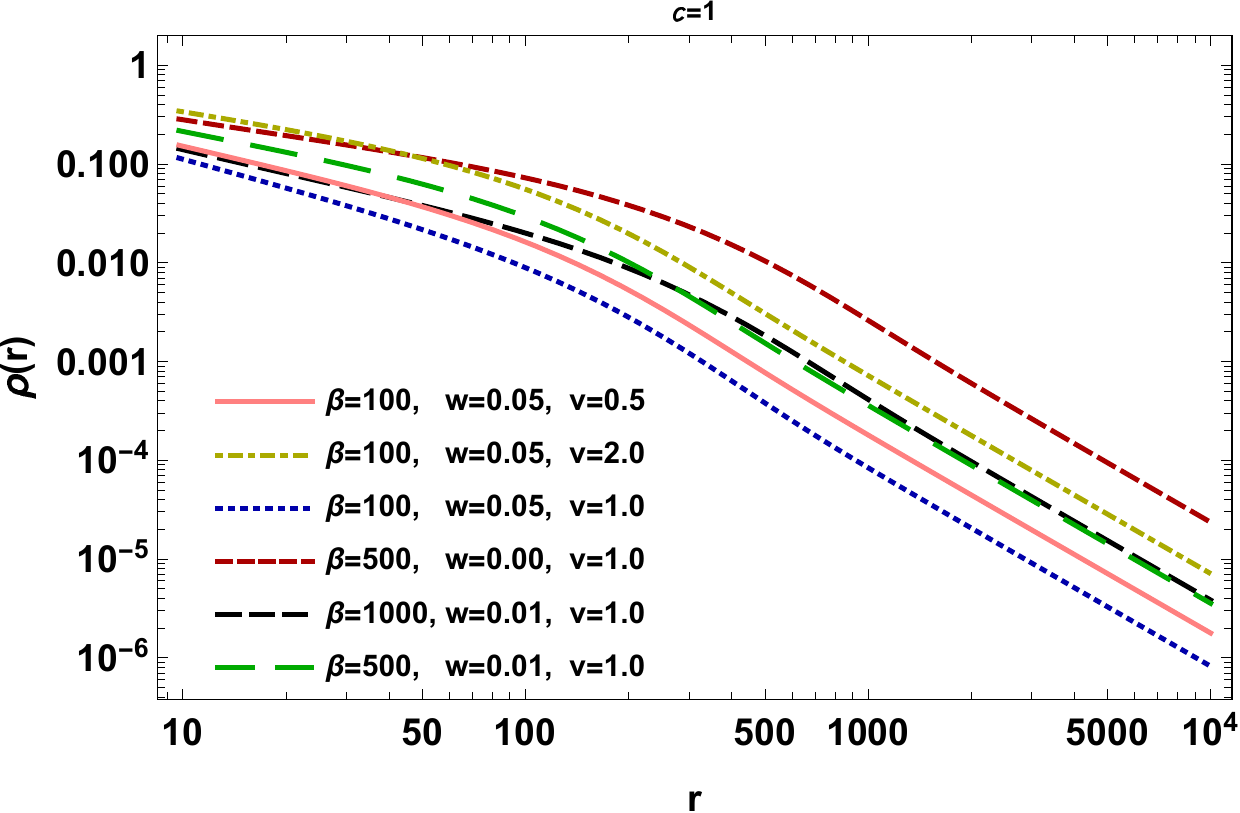,width=8.cm}\hspace{2mm}
\epsfig{figure=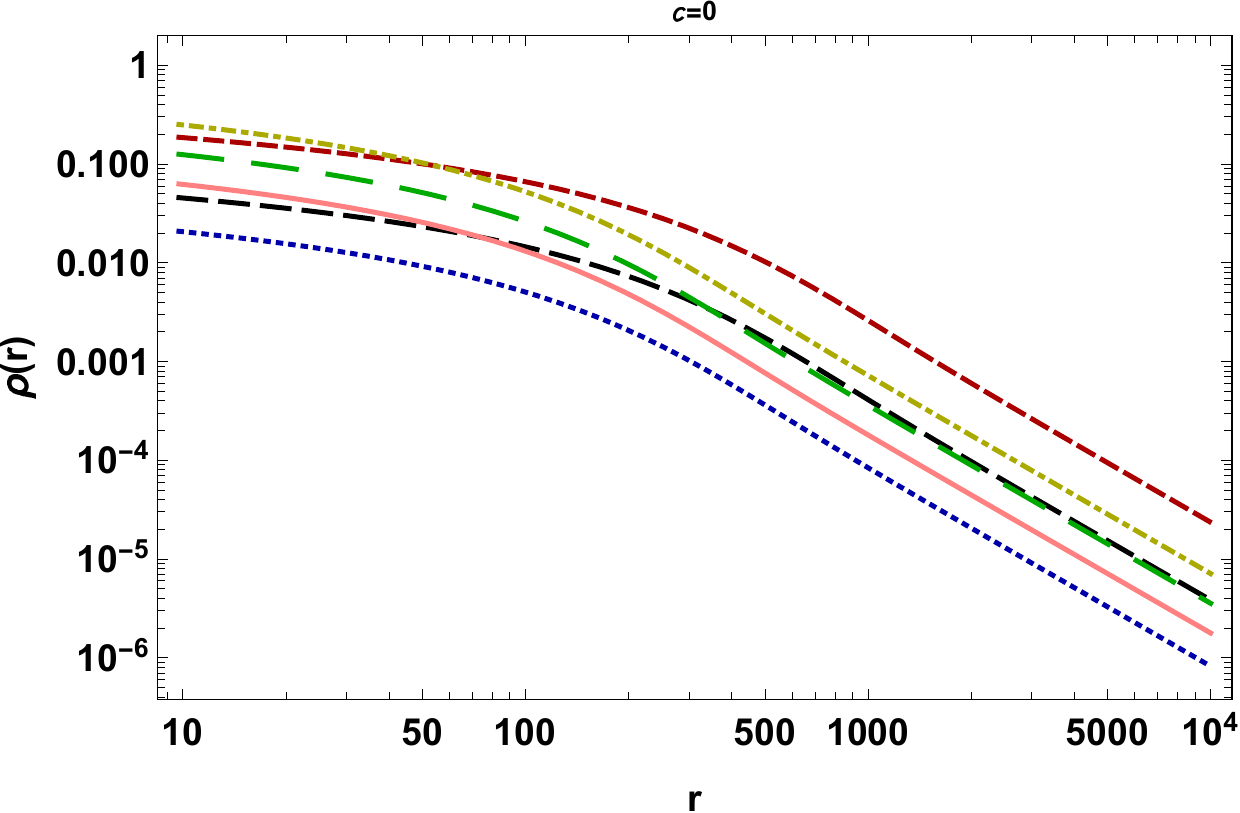,width=8.cm}\vspace{2mm}
\epsfig{figure=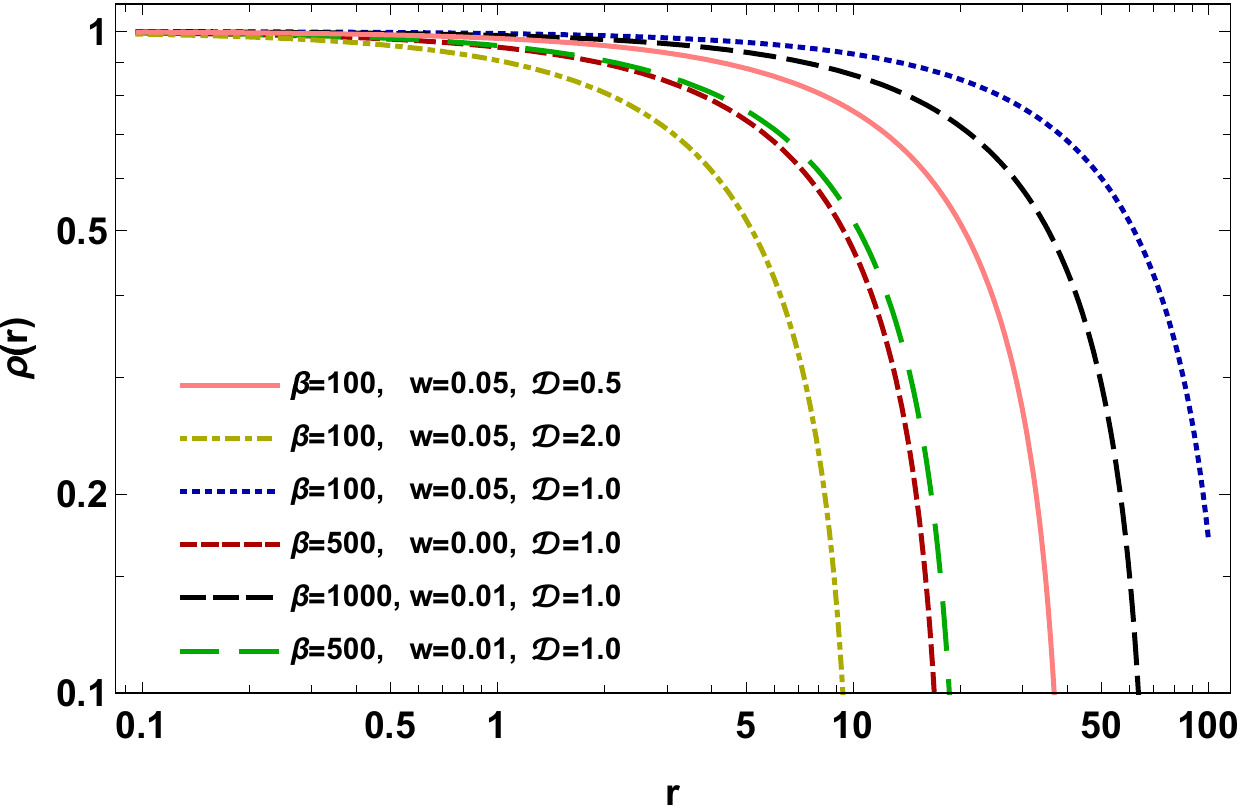,width=8.cm}
\caption{The behavior of the galaxy density in the limit of far and near distances from the center of the galaxy in the context of the $f(R,T)=R(1+\beta \sqrt{T})$ gravity. Upper panels show the solution in far distances for two values of $\mathfrak{c}$. Lower panel indicates the solution in the near distances for $\rho^{s}_{0}=1.$}\label{fig1}
\end{center}
\end{figure}

Another approach is to find the matter density of a spherical mass using a known rotation curve profile. For instance, in \cite{sofue/2013} it has been shown that the rotation curves of the Milky Way can be explained by taking an exponential density distribution of the form $\rho^{ESM}(r)=\rho_{c}\exp{(-r/r_{0})}$, in which $\rho_{c}$ and $r_{0}$ are the central density of the bulge of the galaxy and a scale radius, respectively. The superscript ``ESM'' stands for ``exponential sphere model'', which is the name of this model \cite{sofue/2013}. The mentioned density profile corresponds to the following gravitational potential

\begin{align}\label{eq26}
\Psi^{ESM}(r)=\rho_{c}r^{2}_{0}\left[\frac{1}{2} \left(e^{-\frac{r}{r_{0}}}+1\right)+\frac{r_{0}}{3r}\left(e^{-\frac{r}{r_{0}}}-1\right)\right]+\Psi^{ESM}_{c},
\end{align}
in which $\Psi^{ESM}(0)=\Psi^{ESM}_{c}$. 

Solving Eq.(\ref{eq23}) for the potential~(\ref{eq26}) gives
\begin{align}\label{eq27}
\rho^{f(R,T)}(r)&=\frac{\rho _c e^{-2 r \left(\frac{1}{r_0}+\frac{i \sqrt{\mathcal{D}^b}}{\sqrt{\mathcal{M}^b}}\right)}}{4 r \sqrt{\mathcal{D}^b} \left(r_0^2 \mathcal{D}^b+\mathcal{M}^b\right){}^2}\left\{\sqrt{\mathcal{M}^b} e^{\frac{2 r}{r_0}+\frac{i r \sqrt{\mathcal{D}^b}}{\sqrt{\mathcal{M}^b}}} \left(\sqrt{\mathcal{M}^b}+i r_0 \sqrt{\mathcal{D}^b}\right){}^2 \left[4 r_0 \sqrt{\mathcal{D}^b} \sqrt{\mathcal{M}^b}+i r_0^2 \left(1-2 \mathcal{D}^b\right)+2 i \mathcal{M}^b\right]-\right.\nonumber\\
&i \sqrt{\mathcal{M}^b} e^{\frac{2 r}{r_0}+\frac{3 i r \sqrt{\mathcal{D}^b}}{\sqrt{\mathcal{M}^b}}} \left(\sqrt{\mathcal{M}^b}-i r_0 \sqrt{\mathcal{D}^b}\right){}^2 \left[4 i r_0 \sqrt{\mathcal{D}^b} \sqrt{\mathcal{M}^b}+r_0^2 \left(1-2 \mathcal{D}^b\right)+2 \mathcal{M}^b\right]\nonumber\\
&\left.+2 r_0^2 \sqrt{\mathcal{D}^b} e^{r \left(\frac{1}{r_0}+\frac{2 i \sqrt{\mathcal{D}^b}}{\sqrt{\mathcal{M}^b}}\right)} \left[r r_0^2 \mathcal{D}^b+\left(r+2 r_0\right) \mathcal{M}^b\right]\right\}.
\end{align}

\begin{figure}[h!]
\begin{center}
\epsfig{figure=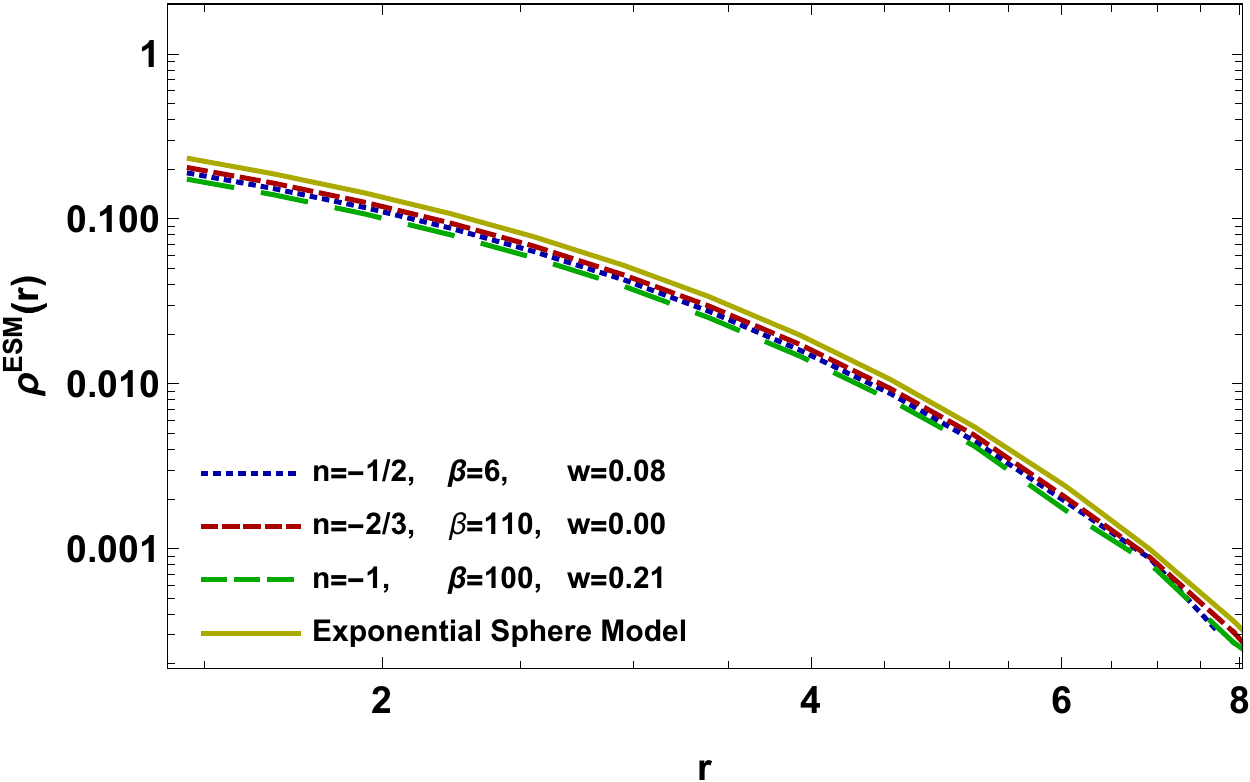,width=8.cm}
\caption{Comparison of the galaxy density profiles of exponential sphere model to the one obtained from $f(R,T)=R(1+\beta \sqrt{T})$ theory. $\rho _c=1=r_{0}$ has been used.}\label{fig2}
\end{center}
\end{figure}

Solution~(\ref{eq27}) leads to the same rotation curves discussed in~\cite{sofue/2013}. Note that, although in this solution the imaginary number $i$ appears, nevertheless, it gives real valued profiles for some amounts of the free parameters $\beta$ and $w$. We compare the solution (\ref{eq27}) with the density profile of exponential sphere model in Fig.~\ref{fig2}. It can be seen that in the framework of non-minimal oupling $f(R,T)$ model, one can tackle the problem of galaxy rotation curves. The phenomenological density profile of the exponential sphere model can be replaced by a completely theoretical solution which does produce the same effect on the galaxy rotation curves.

\section{Conclusion remarks}

Since the seminal works by Rubin and collaborators \cite{rubin/1978}-\cite{rubin/1979}, theoretical physicists have tried hard to explain the flattened pattern that shows up for large distances to the galactic center. The ``standard'' approach is to assume there is a dark matter halo involving the galaxies. 

There is a number of models that theorize dark matter particles, such as axions \cite{duffy/2009} and neutralinos \cite{bednyakov/1997}, among many others. None of these dark matter particles has been experimentally probed so far (check the references in the Introduction section, among others). 

This has led some theoretical physicists to try to explain the rotation curves of galaxies with no dark matter, but by changing the dynamical laws. For example, Milgrom phenomenologically changed the newtonian gravitational law in the limit of low acceleration (check Reference \cite{milgrom/1988} for example, among others), such as the outer regions of spiral galaxies.

Even in the structure formation scenario, some alternatives to dark matter can be seen, as one can check, for instance, \cite{dodelson/2006}-\cite{pal/2006}.   

Returning to the galaxy rotation curves issue, a more fundamental form of treating it with no dark matter is by considering alternative (or modified or extended) theories of gravity in their weak-field regime, as it was done in \cite{deliduman/2020}-\cite{mak/2004} for instance.  

Here we have followed the above approach for the $f(R,T)$ theory of gravity. We highlighted the ability of the $f(R,T)$ gravity theories to describe the flatness problem of galaxy rotation curves. Distribution of a spherical mass immersed in the background cosmic matter has been considered. Our investigation was done on two classes of models: $f(R,T)=R+\alpha T^{n}$ models, which are called ``minimal coupling models" and $f(R,T)=R(1+\beta T^{n})$ models, which are called ``non-minimal coupling model''.

We have shown that the former type of models cannot bring forward an important correction with respect to GR results. In these models, the correction term is proportional to the inverse of distance which implies non-significant deviation from GR. 

To inspect the physical content of the field equations using the non-minimal coupling models, we obtained far and near-distance solutions considering a linear and a constant approximation for the velocity rotation curves, respectively. Correspondingly, two matter density profiles have been obtained with the property of being decreasing functions of distance. We showed that in the $f(R,T)$ gravitational non-minimal coupling model, a local perturbation of a pressureless cosmic matter yields spherical masses with flat rotation curves without demanding the existence of a dark matter halo. 

Also, we achieved the matter density profile for a known velocity profile which is phenomenologically suggested in the Exponential Sphere Model~\cite{sofue/2013} and we demonstrated that our solution behaves the same as the corresponding phenomenological density profile.  \\




\end{document}